# Population III Stars in I Zw 18

Sara Heap (NASA's Goddard Space Flight Center, USA)

Jean-Claude Bouret (Laboratoire d'Astrophysique de Marseille, France)

Ivan Hubeny (University of Arizona, USA)

ABSTRACT

Ultraviolet and 21-cm observations suggest that the extremely low-metallicity galaxy, I Zw 18, is a stream-fed galaxy containing a "pocket" of pristine stars responsible for producing nebular He II recombination emission observed in I Zw18-NW. Far-UV spectra by Hubble/COS and the Far Ultraviolet Spectroscopic Explorer (FUSE) make this suggestion conclusive by demonstrating that the spectrum of I Zw 18-NW shows no metal lines like O VI 1032, 1038 of comparable ionization as the He II recombination emission.

1. INTRODUCTION

I Zw 18 is a blue, compact dwarf galaxy at a distance, D~18 Mpc (Aloisi et al. 2007). As the SDSS image of I Zw 18 (Fig. 1 left) shows, the optical galaxy contains two young, massive star clusters, one to the northwest (NW), the other to the southeast (SE). The metallicity of the surrounding ionized nebula is among the lowest known: log O/H + 12 = 7.18. Because of its extremely low metallicity, I Zw 18 has been touted as the best analogue to primitive galaxies in the early universe, and it has been observed by virtually all the major telescopes in space and on the ground. VLA 21-cm observations (Fig. 1 right) show that the optical galaxy is embedded in an extensive neutral envelope. The H I cloud completely encloses the optical galaxy.

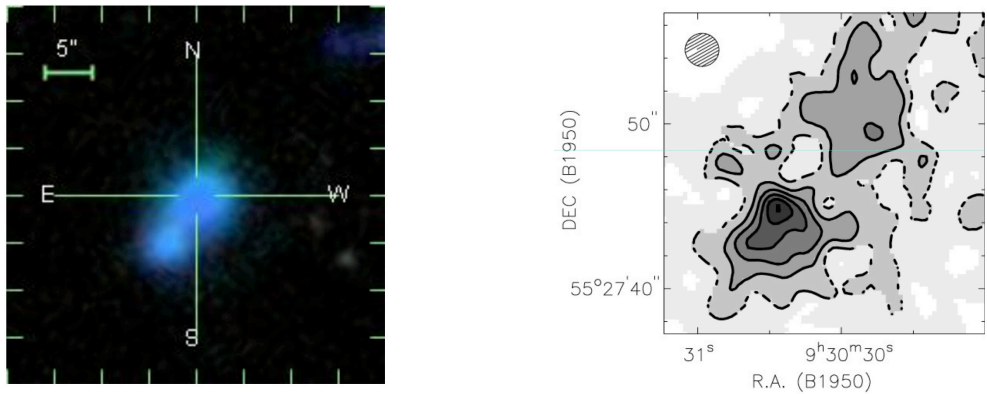

*Fig. 1. Left: SDSS image of I Zw 18; Right: VLA 21-cm map of I Zw 18 [Lelli et al. 2012]*

Below, we contend based on direct and indirect arguments that I Zw 18 hosts a "pocket" of Pop III stars in the NW component.

## 2. INDIRECT ARGUMENTS

There are several observations that *suggest* that I Zw 18 has a population of Pop III stars.

2.1. The metallicity of I Zw 18 is far too low for its luminosity

As pointed out by Skillman et al. (2013), I Zw 18 does not conform to the Luminosity-Metallicity Relation of dwarf galaxies established by Berg et al. (2012). It is either 0.7 dex too low in metallicity or 7 magnitudes too bright. Skillman et al. and Ekta & Chengular (2010) suggest that the abnormally low metallicity of I Zw 18 is due to dilution by inflowing gas of even lower metallicity than that of the H II region.

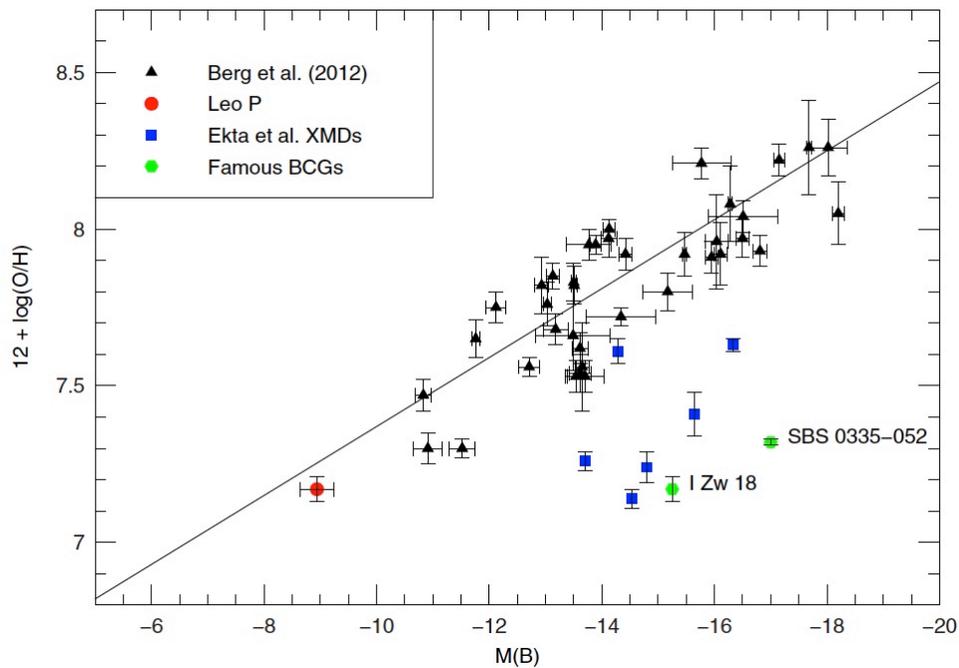

*Figure 2: The abnormally low metallicity of I Zw 18 [Skillman et al. 2013]*

2.2. The neutral envelope of I Zw 18 has a lower metallicity on average than does the ionized region

The neutral gas in IZw 18 is more metal-poor than the ionized gas, with abundances lower by 0.18–0.77 dex depending on the element (Lebouteiller et al. 2013) (Fig 3). The difference in metallicity can result from:

(1) the H II region being enriched by the present or recent star-formation episode(s) or
(2) the presence of extremely low-metallicity clouds in the H I region diluting the observed abundances.

After weighing the various arguments, Lebouteiller et al. conclude that as much as 50% of the H I mass could be pristine.

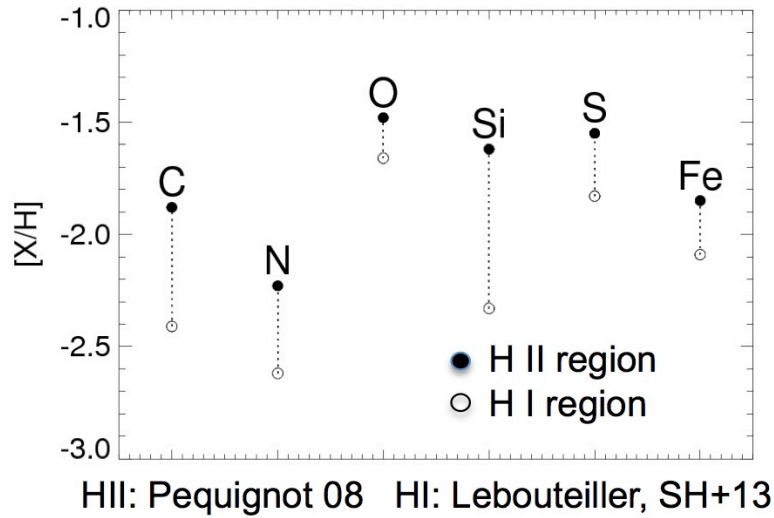

*Figure 3: Metal abundances in the H II and H I regions of I Zw 18.*

2.3. I Zw 18 has a visible inflowing stream akin to predictions for very high redshift galaxies

An accreting stream is visible in the 21-cm map of I Zw 18 (Fig. 4 right). Lelli et al. (2014) find that the main body of I Zw 18 is a rotating disk viewed at a 70$^o$ inclination with the NW side approaching, and the SE side, receding. In addition, a narrow stream is approaching the main body of Zw 18 from the south. According to Lelli et al., this southern stream "does not seem kinematically connected with the SE region of the main body of I Zw 18, as the gas velocity changes abruptly from 790 km s$^{-1}$ to 720 km s$^{-1}$. Moreover, at the junction between I Zw 18 and the stream, the HI line profiles are double peaked, suggesting that there are two distinct components, possibly well separated in space but projected to the same location on the sky".

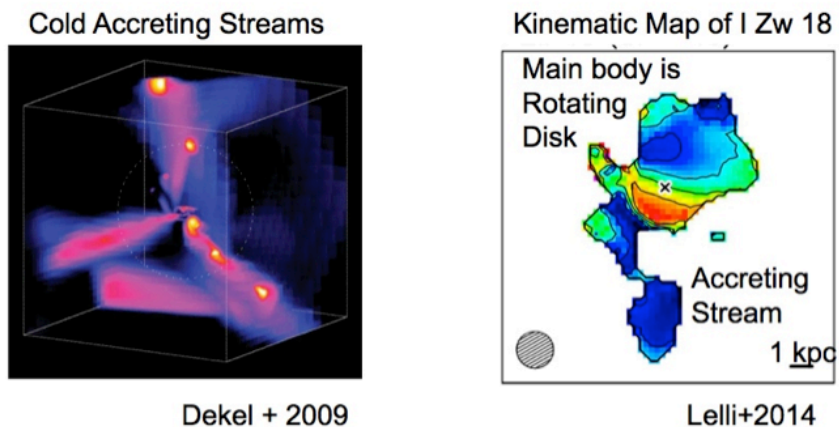

*Figure 4: Left: 3D image of simulations of cold accreting streams (Dekel et al. 2009) Right: Kinematic map of I Zw 18 (Lelli et al. 2014).*

It is likely that the metallicity of the southern stream is well below that of the main body of I Zw 18. It is not clear where the stream makes contact with I Zw 18, but it is likely that some of the stars formed in the collision of the stream with I Zw 18 will be pristine or nearly so.

The inflow to I Zw 18 is suggestive of stream-fed galaxies in the early universe (Fig 4 left). Cosmological simulations by Dekel et al. (2009) predict that massive galaxies in the early universe grew primarily by accretion of narrow streams, which penetrated the dark matter halo and collided with the central galaxy, thereby provoking star formation. This prediction implies that primitive galaxies were already heterogeneous: stars formed in the collision of the stream with the galaxy should be younger and have a lower metallicity reflecting the metallicity of the stream.

3. DIRECT ARGUMENTS

Here, we describe potentially conclusive arguments that I Zw 18-NW hosts metal-free stars.

3.1. The spectrum of I Zw 18-NW shows nebular He II emission implying the presence of extremely hot, luminous stars

Ground-based observers have long noted that I Zw 18-NW shows He II 4686 in emission, and UV spectra obtained with Hubble's Cosmic Origins Spectrograph (COS) show He II 1640 in emission. It is clear that the emission is nebular because (1) the emission lines are too narrow to originate in Wolf-Rayet stars, and (2) the emission is extended (Kehrig et al. 2015). Nebular He II emission results from ionization of He$^+$ by strong stellar He II Lyman continuum radiation and subsequent recombination. Only extremely hot, luminous stars have strong He II Lyman continuum luminosities.

3.2. The ultraviolet spectrum of I Zw 18-NW shows no evidence of extremely hot, luminous stars

Far-UV spectra of I Zw 18-NW obtained with Hubble's Cosmic Origins Spectrograph show that the dominant stellar population is a young, massive OB star cluster. Because winds of OB stars are driven by radiation pressure on metal lines, and their metallicity is so low, these OB stars do not have winds, so a photospheric analysis is sufficient. Figure 5 compares the observed spectra (black) with a model integrated stellar spectrum (red) constructed with non-LTE photospheric code, `synspec` by Lanz & Hubeny (2003, 2007). The model assumes a projected rotational velocity, v sin i =200 km/s and instrumental broadening having a FWHM=0.55 Å. We derive a cluster age of 5 Myr via ionization balance: C IV 1169 vs. C III 1175; O V 1371 vs. O IV 1338, 1342-43. (Note that O V 1371 is not present.). The model produces an acceptable fit to all the spectral lines except C IV 1548,1550 and He II 1640, which are filled by nebular emission. Nebular He II emission is inconsistent with the metal-line spectrum. At an age of 5 Myr, no star is hotter than ~40,000 K. Such a population cannot produce the observed nebular He II emission lines.

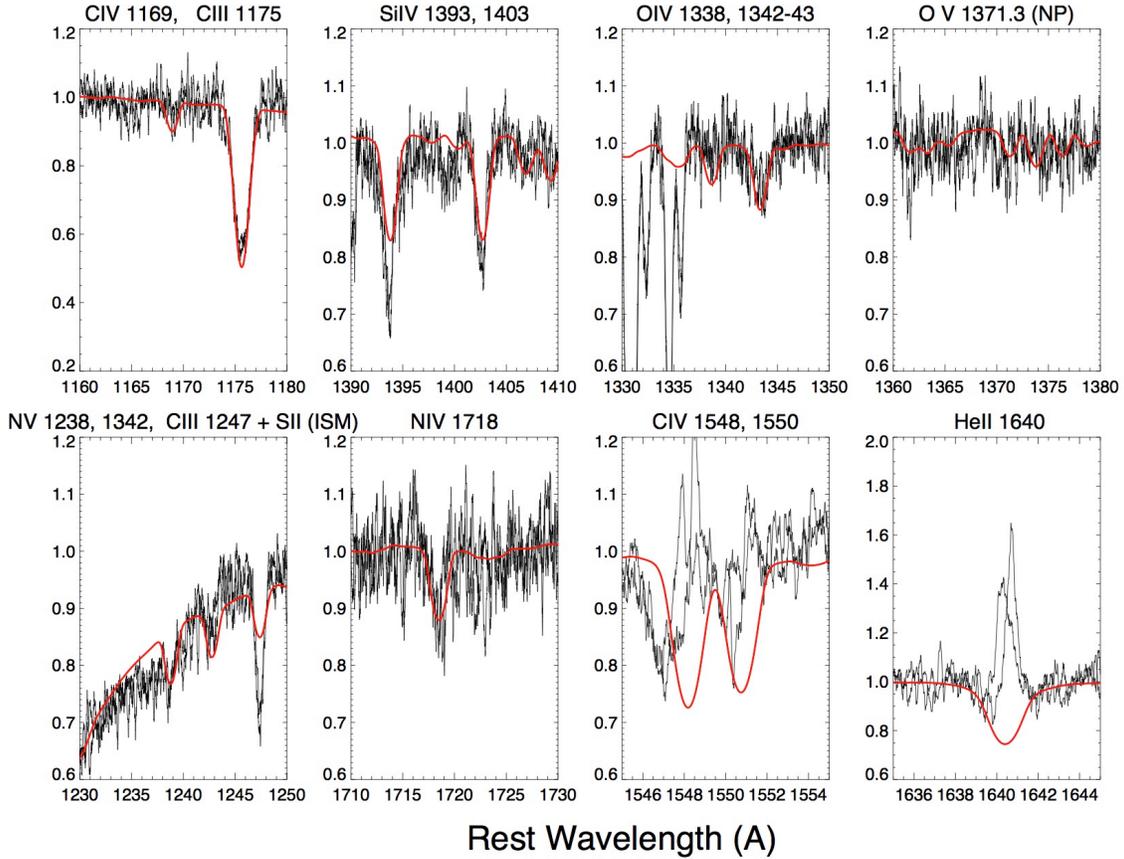

*Figure 5: Comparison of COS spectra (black) with a model spectrum (red) of a 5 Myr old stellar population with metallicity, $Z=1/50\ Z_\odot$.*

The two COS spectra of most features are identical within the noise. However, nebular emission lines, C IV 1548, 1550, He II 1640, and O III] 1661, have different apparent wavelengths in the two spectra, which were taken seven months apart (different orientation of the dispersion axis). The wavelength offsets indicate that the He II-emission source is compact and offset from the center of the 2.5" circular COS aperture. As shown in Fig. 6 Left, it is likely located along the eastern edge of NW star cluster, near or in the bright, compact [O III] knot. This position is consistent with recent IFS observations of He II 4686 by Kehrig et al. 2015 (Fig. 6 Right). We attribute the He II nebular emission to stars formed in the collision of the southern stream with the NW cluster.

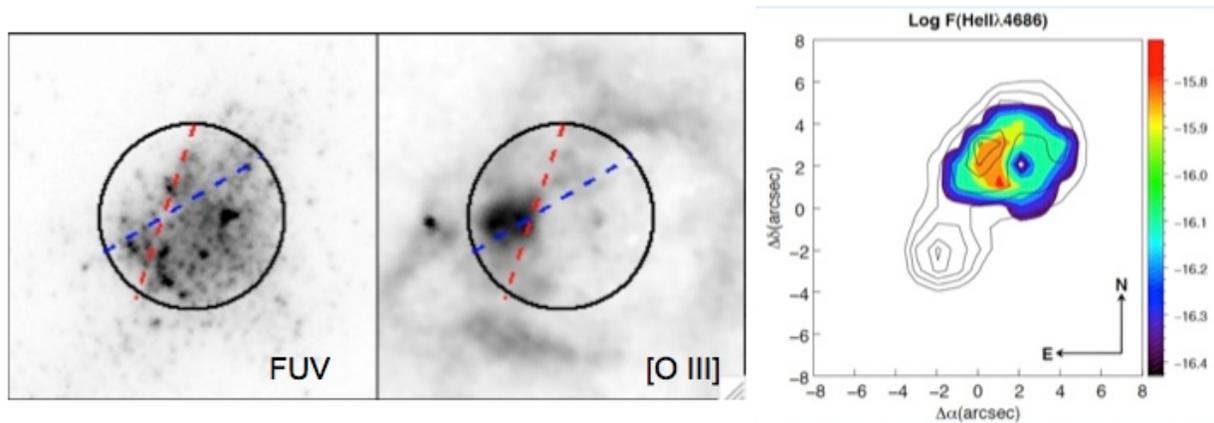

*Figure 6: Left: Position of the source of He emission with respect to the stars (FUV) and nebula ([O III]). The position of the source is at the intersection of the red and blue dashed lines. Right: Map of He II 4686 emission by Kehrig et al. (2015).*

3.3. I Zw 18-NW has two distinct young stellar populations

The two distinct stellar populations in I Zw 18 differ both in ionization level and morphology. The dominant population has an ionization level equivalent to a mid-O star, while the stars responsible for the nebular He II emission are highly ionized, having temperatures of 75,000 K or hotter. The dominant population has the morphology of an open star cluster, while the stars responsible for the nebular He II emission are concentrated in an area along the eastern edge of the star cluster.

We suggest that the OB star cluster is formed by the interaction/merging of the NW and SE components, while the stars producing nebular He II emission were formed by the interaction of the stream with the NW component.

3.4 The stars responsible for the nebular He II emission must be pristine or nearly so

The two stellar populations also differ in metallicity. If the stars responsible for the nebular He II emission had a metallicity like that of the H I region or even 10 times lower, then the observed stellar spectrum would show high-ionization metal lines like O V 1371 and/or O VI 1032, 1038. However, these metal lines are absent in the HST/COS and Far Ultraviolet Spectroscopic Explorer (FUSE) spectra. The only way to reconcile the moderate ionization level of the stellar and nebular metal-line spectrum with the high ionization level of the nebular He II spectrum is for the stars responsible for the nebular He II emission to be metal-free or nearly so.